\documentclass[journal]{IEEEtran}
\usepackage[utf8]{inputenc}
\usepackage[sorting=none, style=ieee]{biblatex}
\usepackage{xcolor, soul}
\sethlcolor{yellow}
\usepackage{float}
\usepackage{enumitem}
\usepackage[pdftex]{graphicx}
\usepackage{multirow}
\usepackage{float}
\usepackage{subcaption}
\usepackage{tabularx}
\usepackage{amsmath}
\usepackage{authblk}
\usepackage{threeparttable}
\usepackage{booktabs, caption, makecell}
\title{COVID-19 in CXR: from Detection and Severity Scoring to Patient Disease Monitoring}
\author[1]{Rula Amer\thanks{Rula Amer
and Maayan Frid-Adar contributed equally to this work}}
\author[1]{Maayan Frid-Adar}
\author[1]{Ophir Gozes}
\author[1]{Jannette Nassar}
\author[1,2]{Hayit Greenspan}
\affil[1]{RADLogics Ltd., Tel Aviv, Israel}
\affil[2]{Department of Biomedical Engineering,
Tel-Aviv University,
Tel-Aviv, Israel}
\date{}

\begin{document}

\maketitle
\begin{abstract}
This work estimates the severity of pneumonia in COVID-19 patients and reports the findings of a longitudinal study of disease progression. It presents a deep learning model for simultaneous detection and localization of pneumonia in chest Xray (CXR) images, which is shown to generalize to COVID-19 pneumonia. The localization maps are utilized to calculate a ``Pneumonia Ratio" which indicates disease severity. The assessment of disease severity serves to build a temporal disease extent profile for hospitalized patients. To validate the model's applicability to the patient monitoring task, we developed a validation strategy which involves a synthesis of Digital Reconstructed Radiographs (DRRs - synthetic Xray) from serial CT scans; we then compared the disease progression profiles that were generated from the DRRs to those that were generated from CT volumes. 
\end{abstract}


\section{Introduction}
\begin{figure}[!t]
  \centering
  \includegraphics[width=0.425\textwidth]{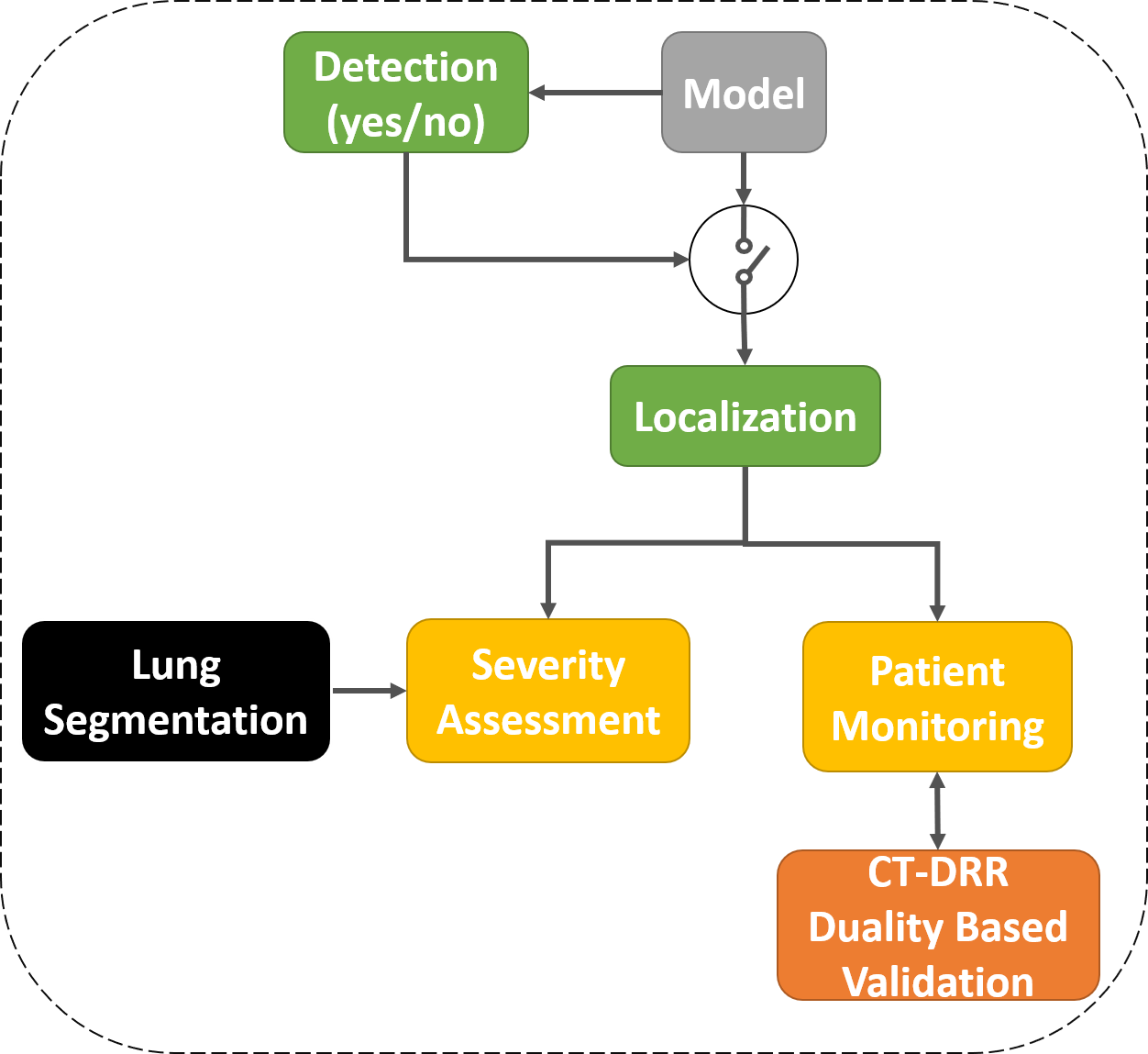}
  \caption{Overview: Detection and localization models are described in Section \ref{section:pneumonia_detection_and_localization}. The methods used for lung segmentation, severity assessment, patient monitoring and CT-DRR duality-based validation are presented in Section \ref{section:severity_scoring}.}
  \label{fig:methods_overview}
\end{figure}

The COVID-19 pandemic is spreading worldwide, infecting millions of people and affecting everyday lives.
Most patients experience mild symptoms including a fever, dry cough, and a sore throat. However, some patients deteriorate and experience complications such as  Acute Respiratory Distress Syndrome (ARDS), organ failure and even death \cite{singhal2020review, chen2020epidemiological, sohrabi2020world}.\\
\indent Studies investigating which imaging modality to use for COVID-19 patients, have compared the advantages of CT vs. Chest Xray (CXR), and vice versa \cite{jacobi2020portable, rubin2020role}. The decision to use one modality over another depends on the phase of the disease and community norms. In countries where access to RT-PCR tests is limited, the general approach is to encourage patients to contact their doctors early. If suspected patients manifest mild symptoms, a CT scan is performed because it is more sensitive to changes in the lungs caused by mild pneumonia than a CXR examination. In contrast, in countries where the directive approach is to instruct patients to wait to go to the hospital until they experience advanced symptoms, the preferred modality is CXR since it clearly shows abnormalities in the lungs. Another factor that favors the CXR is the high contagiousness of the COVID-19 virus. The complications related to patients' transfer CT suites involve the risk of cross-infections along the route, and in the scanning room. In addition there is a lack of sterilization equipment in some parts of the world. These complications therefore favor the use of the CXR modality for the identification and follow-up of COVID-19 patients. CXR is very useful for assessing disease progression in hospitalized patients for whom the disease state is more likely to be advanced.\\
\indent The rapid spread of the coronavirus pandemic has made AI   important to healthcare specialists in terms of the diagnosis and prognosis of the disease. AI is being actively harnessed to fight COVID-19 as shown in recent applications \cite{bullock2020mapping}. Reviews of AI-empowered publications \cite{shi2020review, latif2020leveraging} point to the numerous machine learning-based studies on segmentations of infected regions in CT scans of COVID-19 patients. However, most CXR publications target the classification task for multiple classes \cite{wang2020covid, apostolopoulos2020covid, zhang2020covid, oh2020deep} and provide interpretable and explainable class activation maps (CAM), rather than accurate COVID-19 pneumonia segmentations. Most of these methods were published at the start of the pandemic, and thus trained solely on a few examples of COVID-19 that were mainly aggregated from publications and radiological websites.
In \cite{maguolo2020critic} and \cite{tartaglione2020unveiling}  experiments were conducted to prove that this data selection might cause the network to learn features that are dataset-biased rather than learning disease-specific characteristics, especially when images of different labels are selected from different databases. Since most current works focus on the {\em{diagnosis}} of COVID-19, it is only recently that we see works targeting 
{\em{severity assessment}} of the disease in CXR. Moreover, to the best of our knowledge, almost no AI-based work has studied and validated the follow-up and {\em{patient monitoring}} of COVID-19 patients using chest radiographs.    
\\ 
\indent In this work, we evaluate the degree of severity of pneumonia in COVID-19 patients and monitor patients' disease progression over time. Fig. \ref{fig:methods_overview} illustrates the process flow: We first determine whether pneumonia is present and localize the infected area of the lung (green blocks). Then, by combining a lung segmentation step (black) and applying a threshold over the localization map that produces accurate lesion segmentation, we measure the relative area of the lung that is infected. We then assess the severity of the case as well as monitor patients over time (yellow). For hospitalized patients who have an extended record of disease, we generate a disease profile over time.
To validate our results, we utilize a novel CT-Xray duality (orange). Using the CT and its accurately defined disease extent, we generate a corresponding synthetic Xray, using a newly developed scheme for Digital Reconstructed Radiograph (DRR) generation. Disease profiles in CT and  Xray space are extracted and compared. Another analysis is then conducted to determine the relationship between CT and Xray in defining disease states. 
The detection and localization components of the system are detailed in Section \ref{section:pneumonia_detection_and_localization}. The generation of severity estimates and monitoring in time are presented in Section \ref{section:severity_scoring}. Experiments and results are presented in Section \ref{section:exps and results}, followed by a discussion and conclusion in Section \ref{section:discussion}.
\\
\indent This work makes five main contributions:
\begin{itemize}
  \item We propose a dual-stage training scheme in the detection and localization network, to accurately segment regions in the lungs infected with pneumonia from inaccurate ground truth (GT) bounding boxes. We exploit the Grad-CAM \cite{selvaraju2017grad} algorithm to generate localization proposals, and use them to learn accurate segmentations that are directly outputted from the model.  
  \item We prove our model's ability, when trained on non-COVID-19 pneumonia patients, to generalize the detection, localization, severity scoring, and monitoring of COVID-19 pneumonia cases.
  \item We introduce a robust lung segmentation method, using unconventional augmentation methods such as synthetic radiographs of abnormal lungs, gamma correction, and blob implanting. Our proposed augmentations ameliorate the segmentation of pathological lungs.
  \item We demonstrate the system's ability to measure the spread of pneumonia in the lungs and to track disease progression.
  \item We present a novel validation strategy for the CXR-based patient disease monitoring, by utilizing  CT scans of COVID-19 patients over time, producing corresponding DRRs, and exploring the CT and Xray duality.  
\end{itemize}

\section{Related Work}
Multiple studies have been published on COVID-19 detection in chest radiographs since the outbreak of the pandemic \cite{alom2020covid_mtnet, rajaraman2020iteratively, lv2020cascade, oh2020deep}.
Here we review several related works. For additional reviews, we refer the reader to the overview papers \cite{shi2020review, shoeibi2020automated} and the COVID-specific Special Issues of TMI\footnote{ \url{https://www.embs.org/tmi/}}.

 
 In Wang et al. \cite{wang2020covid}, the  ``COVID-Net" architecture is presented 
 for COVID-19 detection in CXR. Three datasets, collected from different sources \cite{cohen2020covid, Chung2020git, RSNA2019}, were used to train the network to predict three categories: no
infection (normal), non-COVID19 infection, and COVID-19 viral infection. They reported a sensitivity of 0.95, 0.94, 0.91 for each class with a test set of 100 normal, 100 non-COVID-19 pneumonia and 100 COVID-19 images, respectively.
Apostolopoulos et al. \cite{apostolopoulos2020covid} 
adopted state-of-the-art CNNs that were proposed over the last few years for small medical datasets using a transfer learning method. They utilized the public datasets of COVID-19 from \cite{KaggleCovidData, kermany2018identifying} for bacterial pneumonia, viral pneumonia of COVID-19, and normal image classification. The authors reported the results for 10-fold-cross-validation on two datasets of COVID-19, common bacterial pneumonia (with and without non-COVID-19 patients) and normal cases. Optimal results with a sensitivity and specificity exceeding 0.96 were obtained with the MobileNet v2 network on 224 COVID-19 images. Zhang et al. \cite{zhang2020covid} developed a deep anomaly detection model for COVID-19 vs. non-COVID-19 pneumonia classification. They used 100 COVID-19 images from \cite{cohen2020covid} and 1431 additional CXR images confirmed as other pneumonia from the public
ChestX-ray14 dataset \cite{wang2017chestx}. They reported an $AUC$ of 0.95. 

Severity scoring has also attracted increasing attention in CXR publications \cite{signoroni2020end, cohen2020predicting, li2020improvement, el2020end, li2020automated, zebin2020covid}. Signoroni et al. \cite{signoroni2020end} designed a multi-purpose network for COVID-19 pneumonia prediction, lung segmentation and lung alignment that outputs the severity prediction by dividing the lungs into 6 regions. They utilized 5,000 annotated CXR images from the ASST Spedali Civili of Brescia, Italy, in addition to 194 images from the public dataset in \cite{cohen2020covid}. The mean absolute error (MAE) of the severity score on a subset of 150 images from the private dataset was 1.8 compared to the gold standard with a correlation coefficient of 0.85. The MAE on the 194 images from the public dataset was 2.18.
Cohen et al. \cite{cohen2020predicting} developed a model to predict COVID-19 pneumonia severity based on CXR: they pre-trained a DenseNet on 18 common radiological findings from multiple public datasets, and then trained a linear regression model on a subset of the COVID-19 dataset that was scored by three experts to predict the severity scores 
using different sets of extracted features. The correlation coefficient, $R^2$ and MAE, on a test set of 50 images were 0.78, 0.58 and 0.78, respectively for the pneumonia extent score, and 0.8, 0.6 and 1.14, respectively for the opacity score (encounters with the opacity texture features of consolidation/ground glass).

Here we use the severity scoring to evaluate our predictions of the infected lung area, and focus on both COVID-19 detection and severity scoring in CXR to present an end-to-end solution for COVID-19 disease management.


\begin{figure}[!t]
  \centering
  \includegraphics[width=0.49\textwidth]{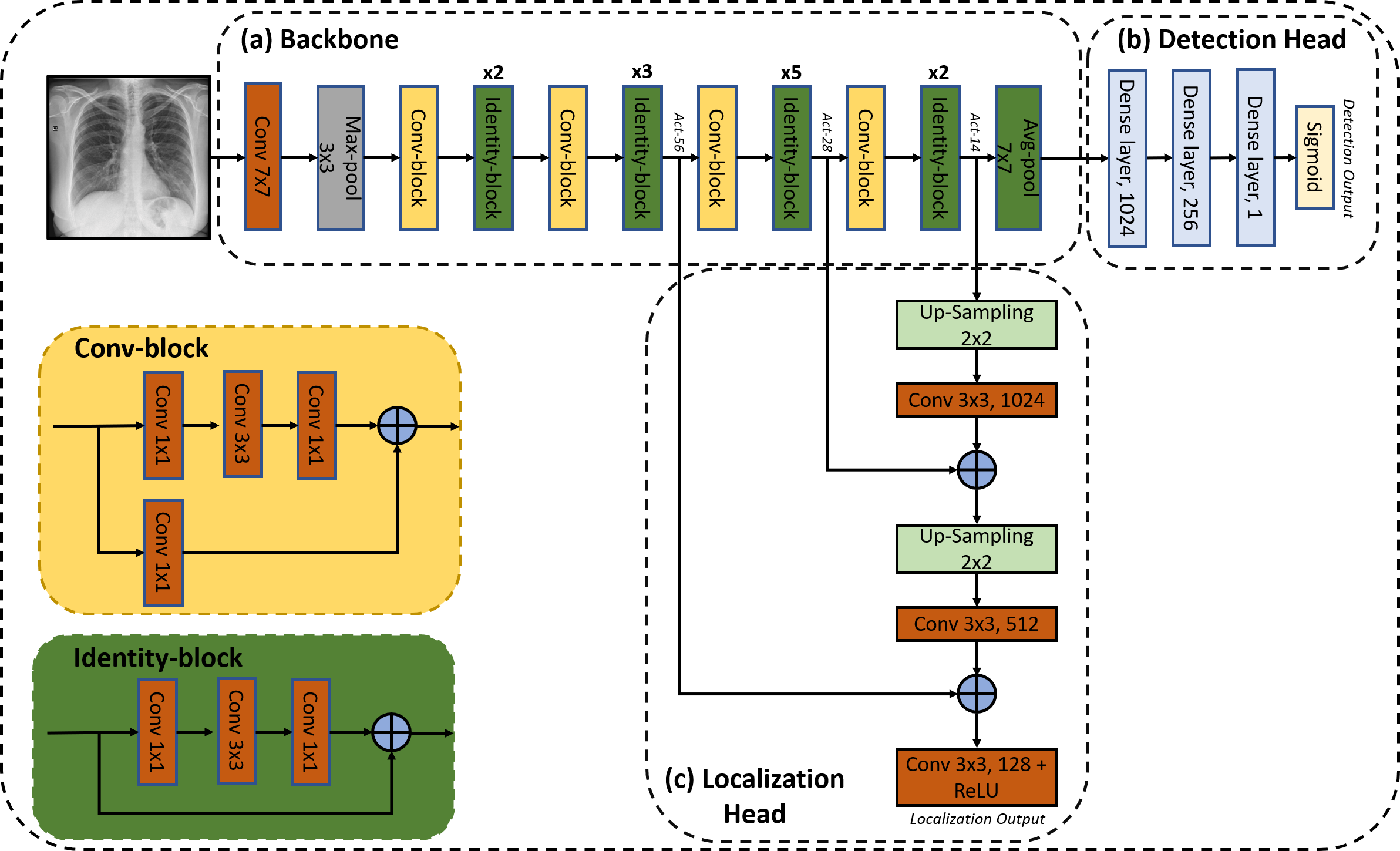}
  \caption{Diagram of the proposed Detection and Localization network. (a) Backbone: pre-trained ResNet50, (b) Detection Head: detection of pneumonia and (c) Localization Head: fuses intermediate convolutional layers of the ResNet50 to form a localization prediction.}
  \label{fig:network}
\end{figure}

\section{COVID-19 Pneumonia Detection and Localization}
\label{section:pneumonia_detection_and_localization}
To assess the severity of pneumonia in COVID-19 patients, the pneumonia region in the CXR of positive patients needs to be accurately segmented. 
In this section we introduce our pneumonia detection and localization network which involves a two-stage training methodology,  to generate fine-grained localization maps from coarse ground truth labels.

\subsection{Detection and Localization Network}
Grad-CAM \cite{selvaraju2017grad} has become a useful tool for localizing COVID-19 pneumonia infection in CXR \cite{zhang2020covid, oh2020deep}. This method is generally used when localization GT data are not available. In this scenario the Grad-CAM method enables only rough localization. Training a network that combines detection and localization would allow for a more accurate disease extent evaluation.

We propose a deep-learning model to predict pneumonia labels and localization maps simultaneously. An illustration of the proposed network is shown in Fig. \ref{fig:network}. It consists of three components: a backbone, a detection head and a localization head. A detailed description of each component is given below.

The backbone is a 50-layer residual network (\textit{ResNet50}) \cite{he2016deep}. 
The network is pre-trained on the ImageNet dataset. 
As shown in Fig. \ref{fig:network}(a), the images are fed to a convolutional layer with $7 \times 7$ kernels and a stride of 2, followed by a $3 \times 3$ max-pooling layer with a stride of 2. This is followed by convolution and identity blocks with skip connections. Each convolution block has 3 convolution layers and another convolution layer in the skip connection, and each identity block also has 3 convolution layers.

The last dense layer of ResNet50 is replaced with three consecutive dense layers with 1024, 256 and 1 neurons, respectively. A dropout layer is inserted between the first two dense layers. Finally, a sigmoid activation function is applied to generate the pneumonia prediction of the detection head.

The localization head is a feature pyramid-like network \cite{lin2017feature}, as shown in Fig. \ref{fig:network}(c).
Low resolution features extracted from the final identity block,
termed $Act-14$, are upsampled by a factor of 2 using nearest neighbor interpolation. The upsampled features undergo a $1 \times 1$ convolution layer to reduce the channel dimensions.
Next, each lateral connection fuses feature maps of the same spatial size from the previous residual block output ($Act-28$) by element-wise addition. This process is repeated for the higher resolution features (activation output of the previous identity block: $Act-56$). Finally, a $3 \times 3$ convolution layer followed by ReLU activation is applied to the last summation and forms the localization output.
The last convolution layer formed has 128 feature maps of size $56 \times 56$, which enables a level of uncertainty in the localization output edges.
In order to create one localization map in the inference stage, the maximal value of the 128 output maps is taken for each matching pixel. 
\subsection{Training Pipeline}
The proposed network is trained using the RSNA Pneumonia Detection Challenge Dataset \cite{RSNA2019}. Pediatric patients were removed from the dataset to prevent bias due to age. The remaining images in the RSNA dataset are split to three sets: training (9004 images), validation (1126 images)  and testing (1124 images). 
The dataset includes annotation labels of pneumonia/non-pneumonia (in equal amounts) and bounding box annotations of the pneumonia regions.
The training images are resized to a fixed size of $448 \times 448$ pixels. A pre-processing step consisting of a Contrast Limited Histogram Equalization (CLAHE) method is applied to the images before training, followed by normalization according to the mean and standard deviation values of the ImageNet database.
\begin{figure}[!t]
  \centering
  \includegraphics[width=0.49\textwidth]{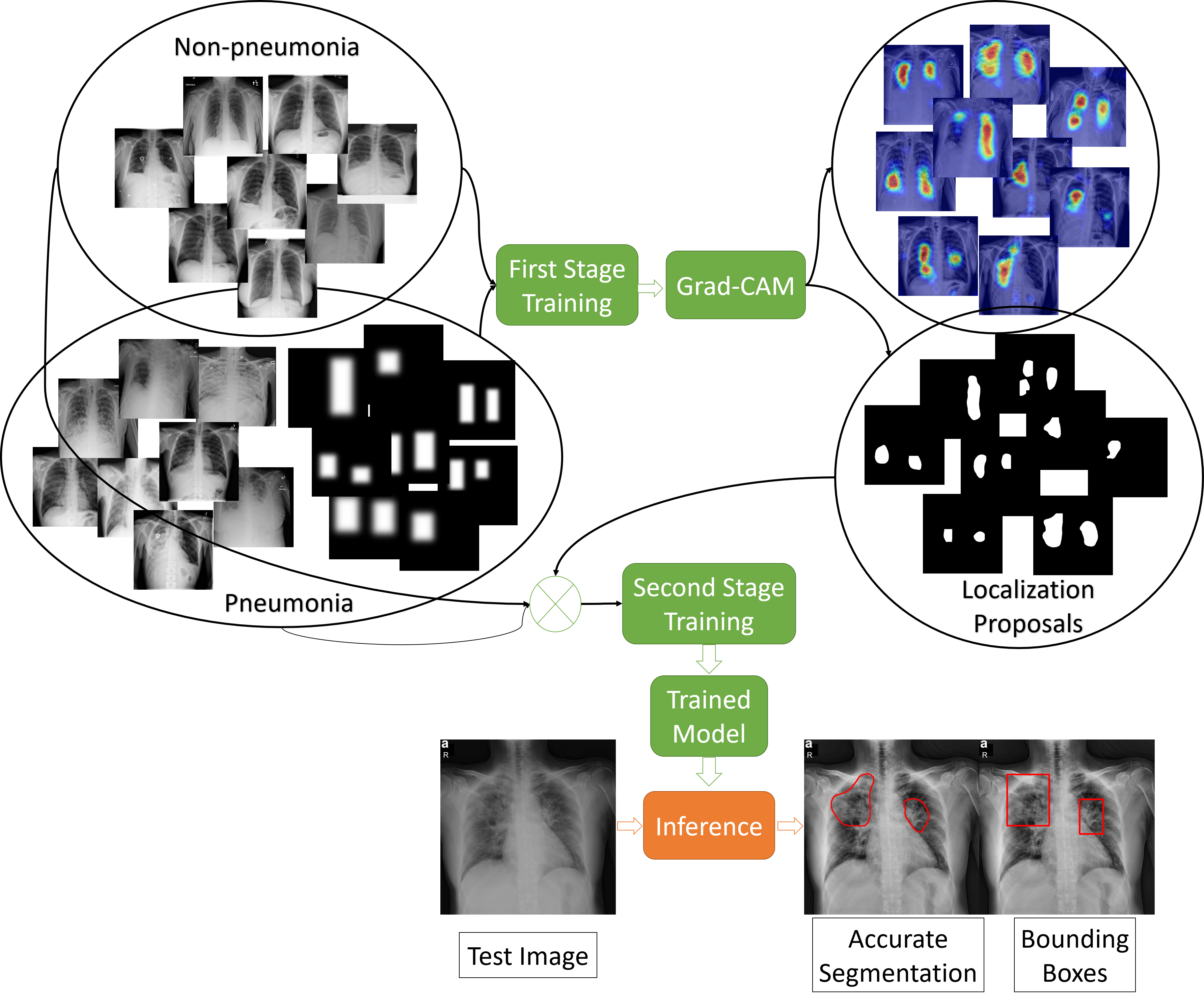}
  \caption{Illustration of the training and testing stages of the Detection and Localization network.}
  \label{fig:training_pipeline}
\end{figure}

\begin{figure}[!t]
  \centering
  \includegraphics[width=0.45\textwidth]{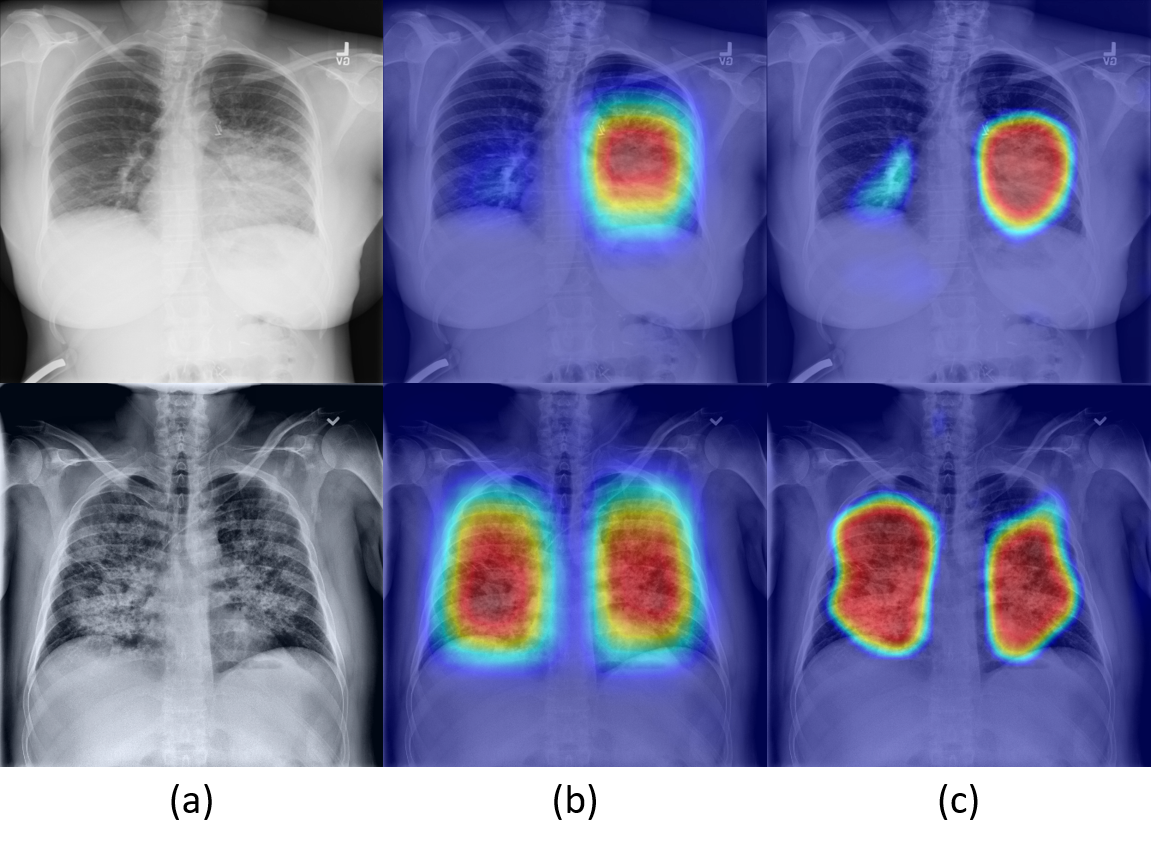}
  \caption{Localization map results of example images as produced from the localization head and presented as heatmap images: (a) input CXR images, (b) heatmaps produced from first stage of the model (DLNet-1), (c) heatmaps produced from the second stage of the model (DLNet-2).}
  \label{fig:pneumonia_heatmaps}
\end{figure}
The training pipeline consists of two stages (see Fig. \ref{fig:training_pipeline}). In the first stage, we train the network on the training images with the corresponding bounding boxes. Next, we use the trained model to generate accurate localization proposals for subsets of the training images using Grad-CAM method, and then replace the bounding box annotations with the accurate localizations to train the model again.
Fig. \ref{fig:pneumonia_heatmaps} compares the produced localization maps of the network after each training stage, and shows the generation of more fine-grained localization after the second stage.
A detailed description of the stages appears below:

\subsubsection{First Stage} 
The network is trained on RSNA data, where the GTs in this stage are the labels of pneumonia/non-pneumonia and the corresponding bounding boxes. We denote this detection and localization model as $DLNet-1$.
Prior to training, a binary image is produced from the bounding boxes where multiple bounding boxes of the same image are combined into one binary image. Then the binary image is dilated with a $5 \times 5$ kernel and finally smoothed with a Gaussian Blur. This last processing step guarantees an accurate prediction of the bounding boxes location, and gives a percentage of uncertainty in the edges. The proposed network is trained using the Adam optimizer. The initial learning rate is set to $1e-4$ and decreases by a factor of 0.2 when learning stagnates for 2 epochs. The batch size is set to 8 images and the max number of epochs
to 30. The loss is comprised of two parts: (1) the detection loss (binary cross entropy) and (2) the localization loss (mean squared error). To compute the localization loss, the localization prediction maps are normalized between 0 to 1 and compared against the binary GT images. The total loss is a linear combination of the two losses, where the binary cross entropy loss on the prediction and the GT label is denoted by $BCE(l_{pred}, l_{gt})$, and the mean squared error by $MSE(BB_{pred}, BB_{gt})$. The total loss is described in (\ref{Eq:total_loss}):
\begin{equation}
Loss = BCE(l_{pred}, l_{gt}) + \lambda MSE(BB_{pred}, BB_{gt})
\label{Eq:total_loss}
\end{equation}
where $\lambda$ is set to $1e-5$ to scale the localization loss according to the detection loss scale.

\subsubsection{Second Stage}
The first trained model is exploited to generate a more accurate pneumonia localization proposals as GT for training the second stage. We denote this model as $DLNet-2$. This is done using the Grad-CAM \cite{selvaraju2017grad} algorithm. Two activation maps are produced, the first generated by back-propagating the gradients from the last convolution layer of the localization head up to the $Act-28$ activation layer, and the second up to the $Act-14$ activation layer. The activation maps are then resized to full image size. The two activation maps are combined to generate one map, whose pixel class probability is more accurate than each map separately. The two activation maps are combined by taking the maximum value of matching pixels from both maps. The final map is then smoothed and normalized. 
To generate the final GT localization proposals,
a threshold of 0.4, a value that gave the highest performance according to intersection with the GT bounding boxes over the testing set, is applied to the fused map.
These accurate localization proposals are multiplied by the GT bounding boxes to eliminate false positives, and then smoothed with a Gaussian Blur to account for possible uncertainties in the edges.
Localization proposals are generated for half of the positive images in the training set that passed a 0.8 detection prediction threshold. The remaining positive images are kept with their corresponding bounding boxes. Those images, together with the negative images, are used for further training the proposed network. The second stage model is trained for 30 epochs with the same training parameters, optimizer and losses that were mentioned in the previous subsection.

\begin{figure}[!t]
  \centering
  \includegraphics[width=0.49\textwidth]{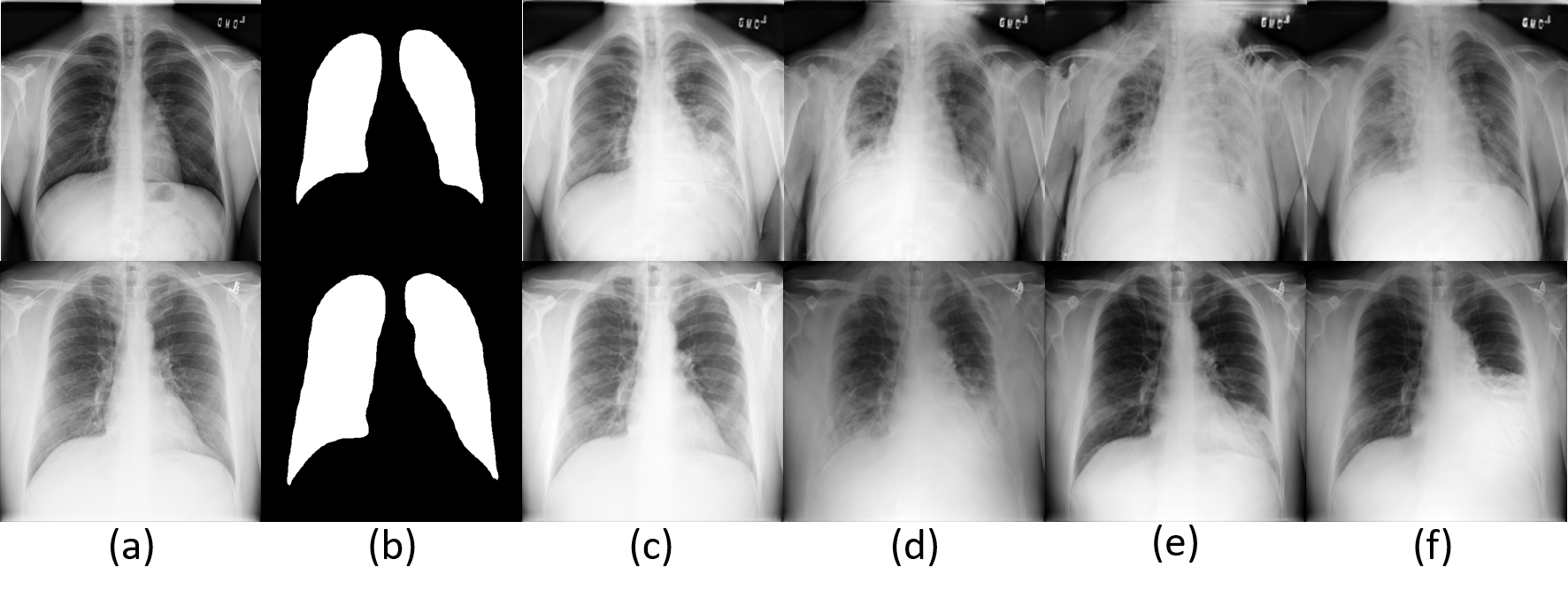}
  \caption{Two examples of  synthesized abnormal CXR images: (a) normal image, (b) corresponding lung segmentation generated by XLSor and (c-f)  abnormal CXRs augmented from the input image  using MUNIT .}
  \label{fig:XLSor_examples}
\end{figure}

\section{Severity Scoring and Patient Monitoring}
\label{section:severity_scoring}
In this section, we focus on measuring the 
extent of pneumonia in the lungs of detected positive patients to assess disease severity. We utilize the severity estimates to monitor patients over time.  A novel validation strategy is proposed that uses the CT-Xray duality: 
we perform validations on digitally reconstructed radiographs (DRRs) synthesized from CT scans and compare them to the original CT images when monitoring the patients' disease state.

\begin{figure*}[!t]
  \centering
  \includegraphics[width=0.75\textwidth]{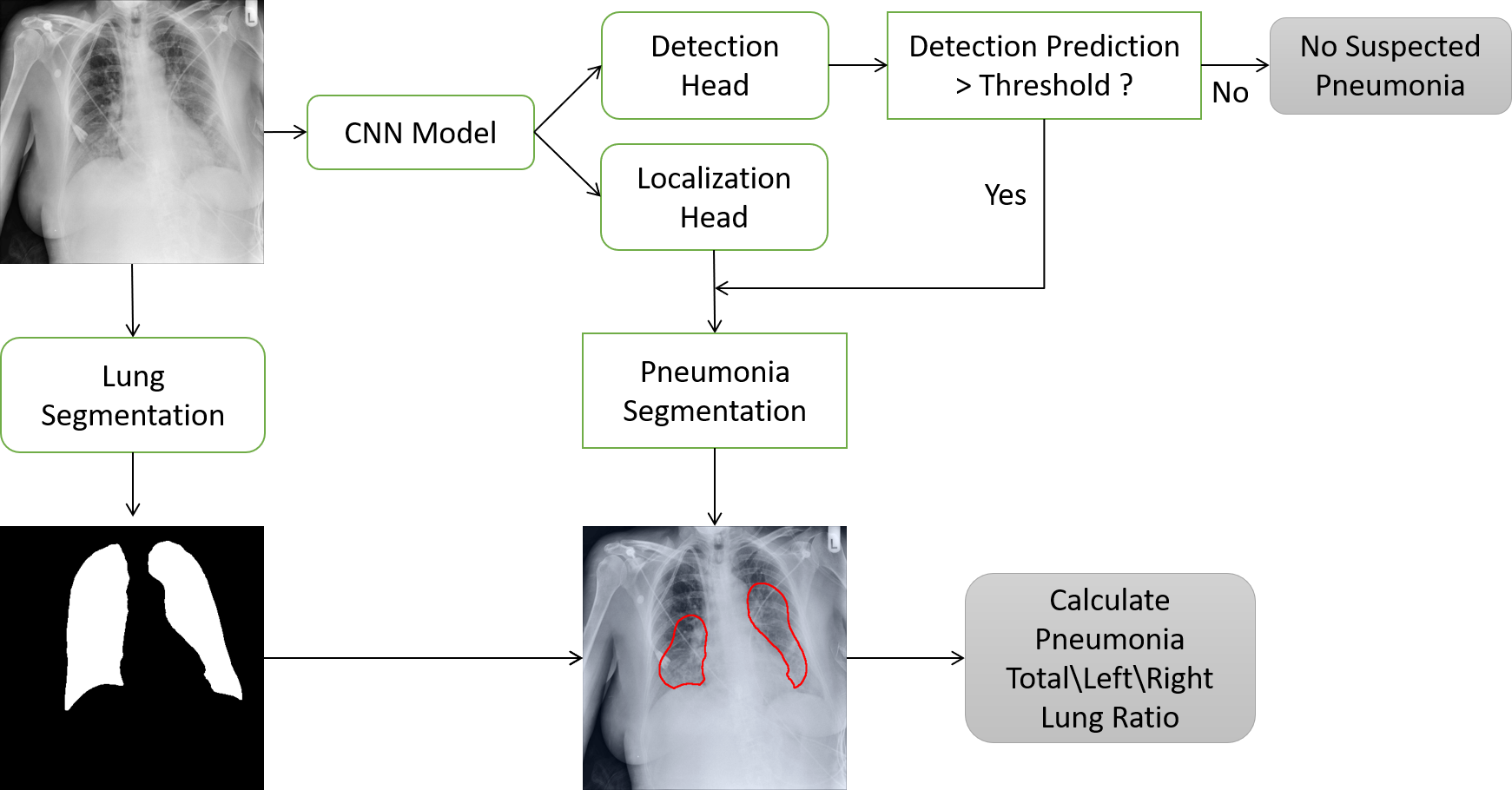}
  \caption{Severity score computation  - Block diagram: The input image enters the detection and localization network.  If the detection prediction is lower than a pre-determined threshold, the image is classified as negative; otherwise, a threshold is applied over the final localization output map to generate the pneumonia segmentation. At this point, the  pneumonia segmentation and the lung segmentation blocks are utilized to compute the ``Pneumonia Ratio".}
  \label{fig:framework}
\end{figure*}

\subsection{Lung Segmentation}
\label{section:lung_segmentation}
To accurately measure the extent of pneumonia in the lungs, we introduce a lung segmentation method for patients with severe opacities and low visibility of the lung fields.

The proposed architecture is a modified U-Net \cite{ronneberger2015u} in which the pre-trained VGG-16 \cite{simonyan2014very} encoder replaced the contracting path (the encoder) in the U-net, as was introduced by Frid-Adar et al. \cite{frid2018improving} for segmentation of anatomical
structures in chest radiographs. The original model, named $LSNet$, was trained on the Japanese Society of Radiological Technology (JSRT) dataset with traditional augmentations (zoom, translation, rotation and horizontal flipping). 
Here, we propose an improved model ($LSNet-Aug$) that is more robust, generalizes to images with severe infections and reduces false detections. The model was improved by challenging the training process with enriched augmented training data. In addition to the original training JSRT dataset, we added images and lung masks GT from the Montgomery County (MC) ‐ Chest X‐ray Database \cite{candemir2013lung, jaeger2013automatic}, the XLSor dataset \cite{tang2019xlsor} and 100 images from the NIH dataset that were provided by the XLSor authors. The XLSor dataset consists of real and synthetic radiographs:  an image-to-image translation module (MUNIT \cite{huang2018multimodal}) is utilized to synthesize radiorealistic abnormal CXRs (synthesized radiographs that appear anatomically realistic) from the source of normal ones, for data augmentation purposes. The lung masks of these synthetic abnormal CXRs are propagated from the segmentation results of their normal counterparts, and then serve as pseudo masks for robust segmentation training.
The aim is to construct a large number of
abnormal CXR pairs with no human intervention, in order to train a powerful, robust and accurate model for CXR lung segmentation. Fig. \ref{fig:XLSor_examples} shows two examples of normal lung images, the GT segmentation maps, and their corresponding synthesized abnormal CXRs.

Additional augmentations were implemented such as gamma correction and blob implanting. The gamma correction simulates Xray images with different intensities from different sources. The blob implanting simulates obstructions in the CXR images, such as tubes, machines and strong infections. The model is trained with Dice loss and optimized using the Adam optimizer. The images are resized to $448 \times 448$ and normalized by their mean and standard deviation. The output score map is thresholded to generate a binary lung segmentation mask.

\subsection{Severity Measurement}
We examined patients that were imaged multiple times during their hospitalization. To evaluate the progression of pneumonia, we suggest a ``Pneumonia Ratio" metric which quantifies the relative area of the segmented pneumonia regions with respect to the total lungs area.

The pneumonia ratio is calculated from both the lung segmentation and the pneumonia segmentation to generate a severity measure of the patient's disease. The lungs are segmented using the lung segmentation module as described above and the segmentation of the suspected pneumonia region is produced by taking the maximal value for each pixel of the 128 predicted localization output maps of the localization head and applying a threshold, 
only for patients that were identified with pneumonia by the detection head. The outputted segmentation map is then multiplied by the lung mask to restrict pneumonia detections to the lung area. The area of the lungs ($Area_{lungs}$) and the pneumonia segmentation ($Area_{pneumonia}$) are calculated according to the total number of pixels involved, and a pneumonia ratio is calculated using the following equation:

\begin{equation}
    Pneumonia\ Ratio = 100 \times \frac{Area_{pneumonia}}{Area_{lungs}}
\end{equation}

The system's components and pneumonia ratio calculation steps are shown in Fig. \ref{fig:framework}.

\subsection{CT and Xray Duality for Patient Monitoring}
\label{subsection:DRR}
To illustrate the efficacy of our model in performing a follow-up task, we describe a strategy to evaluate the accuracy of disease progression using CXR. Rendering realistic DRRs from serial COVID-19 patients' CT scans is manipulated to validate our method. In particular, the $DeepDRR$ framework \cite{unberath2018deepdrr, DeepDRR2019} is implemented to generate DRRs from CT. These DRRs are then inputted to our model to calculate the pneumonia ratio following the steps in Fig. \ref{fig:framework}. The CXR pneumonia ratio is then compared with the CT pneumonia ratio, using the the CT disease localization method described in \cite{gozes2020rapid}.

\begin{figure}[!t]
  \centering
  \includegraphics[width=\columnwidth]{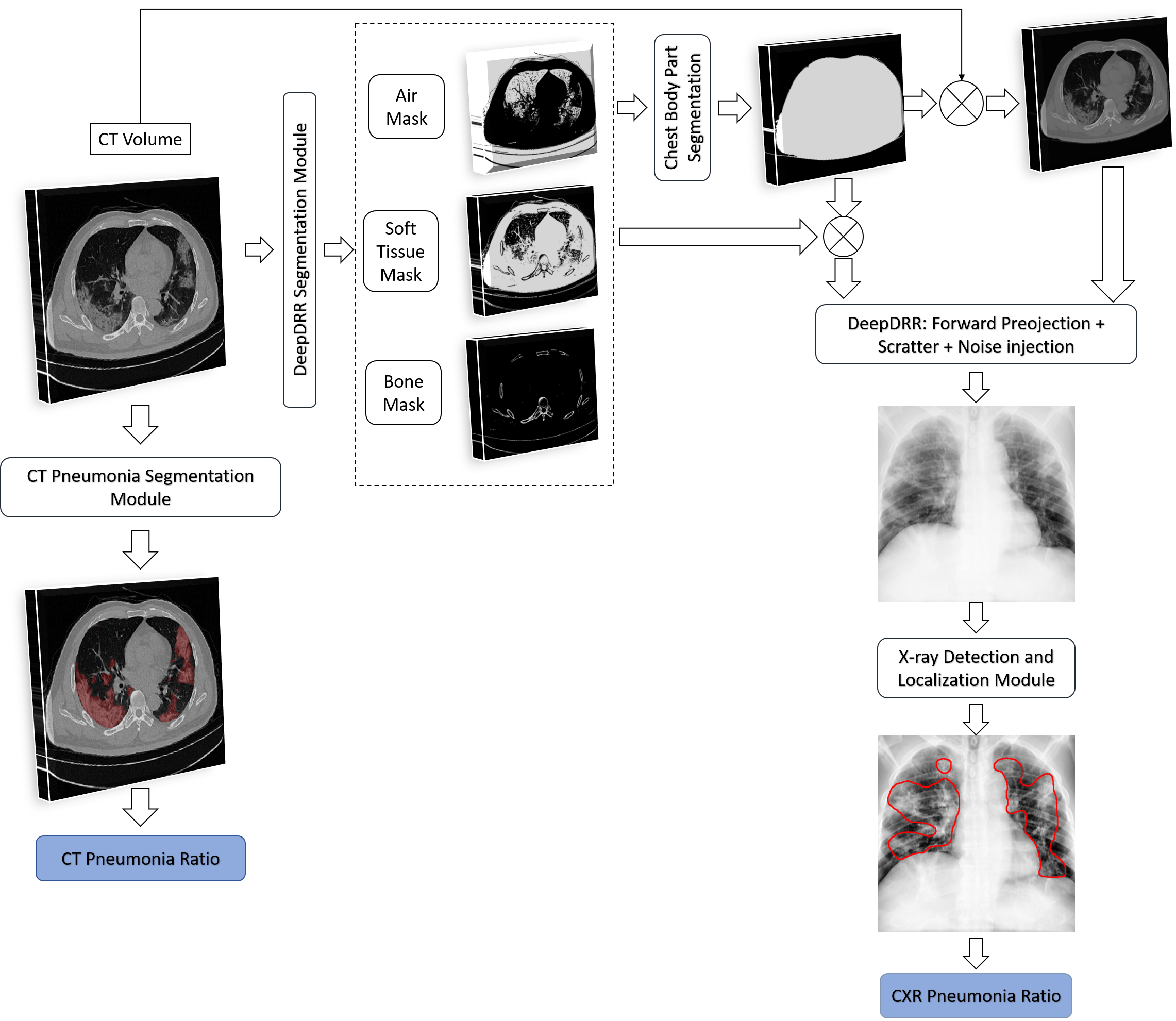}
  \caption{CT and Xray Duality for Patient monitoring. The block diagram shows the steps used to create the DRR; the Pneumonia Ratio can then be computed on both the CT image as well as on the generated synthetic CXR image.  
  }
  \label{fig:DRR_Block_diagram}
\end{figure}

The system of DRR generation and evaluation is depicted in Fig. \ref{fig:DRR_Block_diagram} and described in detail next.
\\
\indent \textit{\textbf{DeepDRR.}} $DeepDRR$ is a machine learning-based method that consists of four modules: (1) material decomposition (air, soft tissue and bones) in CT volumes using a deep segmentation ConvNet, (2) analytic forward projection, (3) scattering estimation in 2D images using a neural network-based Rayleigh, and (4) noise injection. This framework enables the user to generate synthetic Xray images with different parameter configurations, while controlling for image size, resolution, spectrum energy level, image view (rotation), noise and scatter control and others. This can be exploited for data augmentation and parameter tuning. The selected parameters set for the generated DRRs in this work were a $1024 \times 1024$ image size, with a $0.168 \ mm$ pixel size. The spectrum of a tungsten anode operating at $120 \ kV$ with $4.3 \ mm$
aluminum was used and a high-dose acquisition was assumed with $10^5$ photons per pixel. Posterior-anterior (PA) and anterior-posterior (AP) images are produced for each CT volume.\\
\indent \textit{\textbf{Chest Body Part Segmentation.}} A thoracic CT may include scanned objects exterior to the body part such as the patient's bed. These objects are seen on the DRRs, conceal parts of the chest and appear as undesirable noise. Thus, a pre-processing step is applied to keep only the chest parts. First, bit-wise operations are applied to the masks of the decomposed materials: the air mask is inverted using a NOT operation, then an OR operation is performed on the inverted air mask, the soft tissue mask and the bones mask. This step creates a mask of the chest part (without the air in the lungs) as the bed and other unrelated objects are composed of different materials. To produce a binary mask of the whole chest part including the lungs, a hole-fill algorithm is applied. Finally, the filled mask is multiplied by the CT volume, excluding all the unrelated objects.\\
\indent \textit{\textbf{Post-processing.}} The DRRs are first inverted since they appear dark. They are then converted to 8-bit values. Images that are very bright (with an average intensity value exceeding 220) undergo gamma correction with $\gamma=0.2$.

\section{Experiments and Results}
\label{section:exps and results}

\subsection{Datasets}
\label{section:datasets}

To {\em{train our network}}, the main source of data was the RSNA Pneumonia Detection Challenge \cite{RSNA2019, Stein2018} 
These data are comprised of AP and PA and include: $20,672$ radiographs that are labeled  'Normal' or 'No Lung Opacity / Not Normal' indicating that the image is negative for pneumonia, and $6,012$ which are labeled with suspected pneumonia ('Lung Opacity'). The patients in this study ranged in age from $1-100$.

In {\em{testing}} the proposed system, three testing scenarios were used. 
In what we term $Dataset\ 1$, data were set aside from within the RSNA Pneumonia Detection dataset for patients above age 18: 562 CXR images from pneumonia patients, and 562 CXR images diagnosed as healthy or with lung pathologies other than pneumonia (total of 1124 images); the number of $PA$ and $AP$ images was 470 and 654, respectively.

\begin{table}[t]
\caption{Number of images used for training, validation and testing for each dataset.}
\label{table:datasets_details}
\centering
\resizebox{\columnwidth}{!}{
 \begin{tabular}{ccccc} \toprule
 $Dataset$ & $Source$ & $Train$ & $Validation$ & $Test$ \\ \midrule
  Dataset $1$ &  RSNA Pneumonia  
   & 9004 & 1126 & 1124 \\ 
  & Detection Challenge \cite{RSNA2019} & & & \\ \midrule
  Dataset $2$ & COVID-19 Image  & --- & --- & 574\\ 
  &Data Collection \cite{cohen2020covid} & & & \\ 
  &+ RSNA Pneumonia & & & \\
  & Detection Challenge \cite{RSNA2019} & & & \\ \midrule
  Dataset $3$ & COVID-19 Image  & 13604 & 1278 & 300 \\
  &Data Collection \cite{cohen2020covid} & & & \\
  &+ RSNA Pneumonia & & & \\
  & Detection Challenge \cite{RSNA2019} & & & \\
    &+ Figure 1 COVID19 Chest & & &\\
    &Xray Dataset Initiative \cite{Chung2020git}  & & & \\ \bottomrule
 \end{tabular}}
\end{table}

\begin{figure*}[!t]
\begin{subfigure}[b]{0.5\textwidth}
    \includegraphics[width=0.9\textwidth]{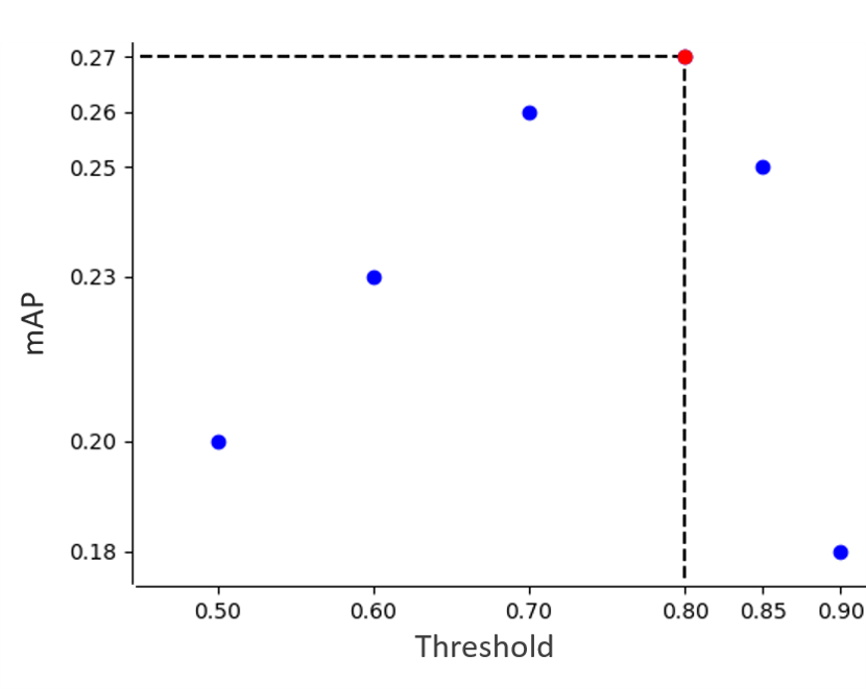}
    \caption{}
    \label{fig:mAP_roc}
  \end{subfigure}
  \begin{subfigure}[b]{0.5\textwidth}
    \includegraphics[width=0.9\textwidth]{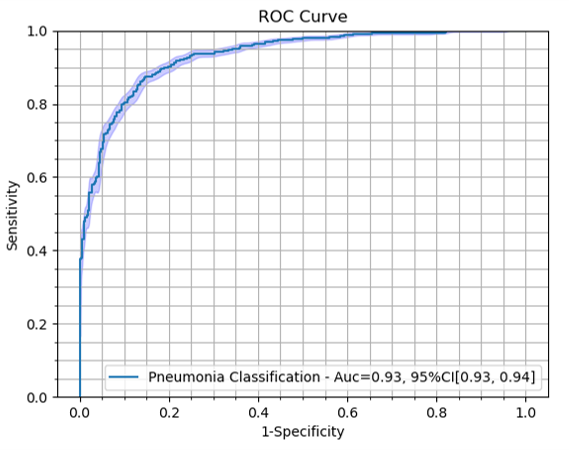}
    \caption{}
    \label{fig:pneumonia_detection_roc}
  \end{subfigure}
  \caption{(a) Mean average precision (mAP) at different thresholds over the localization output of the network. The localization threshold that yields the maximum mAP is selected to produce the segmentations for the final model. (b) ROC curve of our model's performance on pneumonia  detection.}
\end{figure*}

In the second testing scenario, termed $Dataset\ 2$, two data sources were merged:  the main source of the data was the open source COVID-19 Image Data Collection \cite{cohen2020covid}. This dataset consists of COVID-19 cases (as well as SARS and MERS cases) with annotated CXR and CT images; data were collected from public sources as well as through indirect collections from hospitals and physicians. At the time of the writing of this paper, the number of CXR images in the dataset was 339, of which 287 (from 180 patient) $PA$ and $AP$ images and the rest are lateral view position. To balance the data, we randomly selected, and merged, 287 non-pneumonia images from  the RSNA Dataset.
Subsets of $Dataset\ 2$ included additional GT labels, such as lung mask images and severity scoring (see section \ref{COVID-19 Severity Scoring Section}).

Motivated by the COVID-Net experiment conducted in \cite{wang2020covid}, we collected the same dataset and data split for our third testing scenario ($Dataset\ 3$).
This dataset is composed of a total of 8,066 patient cases who have no pneumonia (i.e., normal), 5,538 patient cases who have non-COVID19 pneumonia, and 358 CXR images from 266 COVID-19 patient cases. Of these, 100 normal, 100 pneumonia, and 100 COVID-19 images were randomly selected for testing. A detailed description of the data split for all the datasets used in this paper is shown in Table \ref{table:datasets_details}.

\subsection{COVID-19 Pneumonia Detection}
\label{section:pneumonia_detection}
Several experiments were conducted to evaluate the system's detection performance. Rows $1-5$ in Table \ref{table:detection_and_severity_results} 
summarizes the results over the three datasets defined above in terms of area under the ROC curve ($AUC$), accuracy ($ACC$), positive predictive value ($PPV$), sensitivity ($Sens$) and  specificity ($Spec$).




\begin{table*}[t]
\resizebox{\linewidth}{!}{\begin{threeparttable}
\centering
\caption{Rows $1-5$: Quantitative results of pneumonia detection in COVID-19 and pneumonia patients over three datasets. Rows $6-9$: Comparison of the severity scoring performance metrics of our method after the first/second stage with/without lung segmentation improvement.}
\label{table:detection_and_severity_results}
\begin{tabular}{clccccccc} \toprule

    $Nº$ & $Experiment\ (Detection)$ &  $AUC$ &  $Accuracy$ & $PPV$ & $Sensitivity$ & $Specificity$ \\ \midrule
    1 & Dataset $1$ & 0.93 & 0.86 & 0.86 & 0.87 & 0.85  \\
    2 & Dataset $2$ & 0.94 & 0.89 & 0.90 & 0.86 & 0.91  \\
    3 &  Dataset $3$ & 0.98 & 0.94 & 0.98 & 0.92 & 0.97  \\
    4 &  Dataset $3$ \cite{wang2020covid} & --- & 0.95 & 0.95 & 0.95 & 0.95\\
    5 &   Dataset $3$ using ResNet50\tnote{*} & --- & 0.91 & 0.88 & 0.98 & 0.80  \\ \midrule
    
    & $Method\ (Severity\ Scoring)$ & $Correlation\ Coefficient$ & ${R^2}$ & &\\ \midrule
    
    6 & DLNet-1 + LSNet & 0.75 & 0.38  & & \\
    7 & DLNet-2 + LSNet & 0.79 & 0.59  & & \\
    8 & DLNet-2 + LSNet-Aug & 0.83 & 0.67 & & \\
    9 & Cohen et al. \cite{cohen2020predicting}\tnote{**} & 0.80 & 0.60 & & \\ \bottomrule
\end{tabular}
\begin{tablenotes}\footnotesize
\item[*] Results are taken from \cite{wang2020covid}
\item[**] Results are reported only for 50 test images of \textit{Dataset 2}, the remaining were used for training.
\end{tablenotes}
\end{threeparttable}}
\end{table*}

In the first experiment we evaluated the model's performance for both detection and localization on $Dataset\ 1$, which does not include COVID-19 patients. 
Starting with the evaluation of the pneumonia localization maps (examples over the test set are shown in Fig. \ref{fig:pneumonia_heatmaps}), we measured our proposed network localization predictions vs. GT labels of bounding boxes using an intersection performance metric. For a fair comparison, we used thresholding over the localization map (``localization threshold"), and set a tight bounding box around the segmented region. Different localization threshold values affected the localization performance.
The overall localization performance was assessed by the mean average precision (mAP) at multiple intersection over union ($IoU$) thresholds (``$IoU$ threshold") as suggested by the RSNA pneumonia challenge\footnote{ \url{https://www.kaggle.com/c/rsna-pneumonia-detection-challenge/overview/evaluation}}. The $IoU$ was calculated using (\ref{Eq:IoU}):

\begin{equation}
\begin{array}{l}
IoU(A,B) = \frac{A \cap B}{A \cup B}, \\
    A - prediction,\ B - ground \ truth
\end{array}
\label{Eq:IoU}
\end{equation}

We used $IoU$ threshold values  
from 0.4 to 0.75 with a step size of 0.05, and counted the number of true positive ($TP$), false negative ($FN$), and false positive ($FP$) detections calculated from the comparison of the predicted to the GT  bounding boxes. The suggested precision by the challenge ($RSNA-precision$) of a single image $i$ was calculated at each $IoU$ threshold $t$:

\begin{equation}
RSNA-precision_{i}(t) = \frac{TP(t)}{TP(t)+FP(t)+FN(t)}
\label{Eq:precision}
\end{equation}

The average precision of a single image was calculated as the mean of the above precision values for all $IoU$ thresholds. The overall mAP was then defined as the average of the precision for all the images $i$:

\begin{equation}
mAP = \frac{1}{|images|}\sum_{i} \frac{1}{|thresholds|} \sum_{t} precision_{i}(t)
\label{Eq:mAP}
\end{equation}

The $|images|$ and $|thresholds|$ in the equation indicate the number of the images and $IoU$ thresholds, respectively. In order to define the best localization threshold value over the localization maps that optimized the mAP, we measured the mAP at different threshold values from 0.5 to 0.9 as depicted in Fig. \ref{fig:mAP_roc}. The optimal localization threshold value was 0.8 which resulted in a mAP of 0.27.

\begin{figure}[!t]
  \centering
  \includegraphics[width=0.49\textwidth]{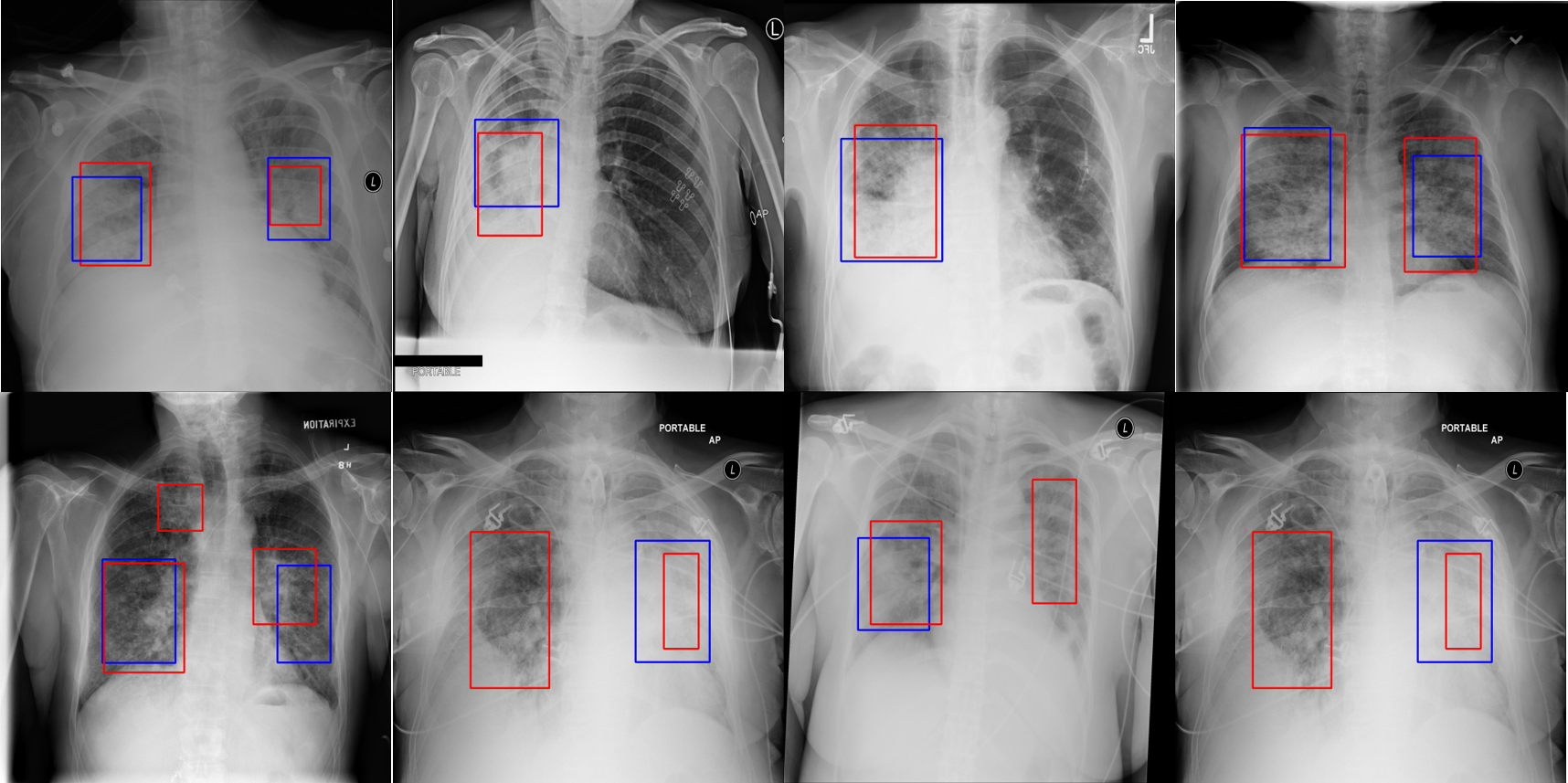}
  \caption{Example results on the test set. The top row depicts successful predictions and the bottom row shows errors. Predictions and GT are shown as red and blue overlays, respectively.}
  \label{fig:bounding_boxes}
\end{figure}

In Fig. \ref{fig:bounding_boxes}, we provide examples of bounding box predictions of our network generated by thresholding over the localization map and set a tight bounding boxes around the segmented regions, in comparison to the GT bounding boxes from the same test set. The top row shows successful predictions and the bottom row depicts discrepancies between the GT and the prediction boxes.

The pneumonia detection performance was evaluated using pneumonia/non-pneumonia labels from $Dataset\ 1$. Fig. \ref{fig:pneumonia_detection_roc} shows the Receiver Operating Characteristic ($ROC$) curve that plots the trade off between sensitivity and specificity at different thresholds on the test set. The reported $AUC$ was 0.93. The $Sens$, $Spec$ and $ACC$ at the optimal predictions threshold of 0.62 was set as the point that satisfied the minimal Euclidean distance from the point (1, 0), are 0.87, 0.85 and 0.86, respectively. The $PPV$, which is the probability that the disease is present when the test is positive was 0.86.



In the second experiment, we examined the model's robustness to COVID-19 data by testing it on $Dataset\ 2$, which includes COVID-19 patients.
The reported $AUC$ for $Dataset\ 2$ is 0.94. The $Sens$, $Spec$, $ACC$ and $PPV$ were 0.86, 0.91, 0.89 and 0.90, respectively, which shows the model's successful generalization to COVID-19 patients' data.

In the last experiment, we use $Dataset\ 3$, which includes pneumonia and COVID-19 patients. For a fair comparison, we trained our network according to the data-split suggested by the authors in \cite{wang2020covid}, where we merged the non-COVID-19 pneumonia and COVID-19 pneumonia images into one class to fit our network. 
In this detection task, we achieved a 0.98 $AUC$, with 0.92 $Sens$ and 0.97 $Spec$. These results are comparable to state-of-the-art performance. When comparing the detection performance of a single network (ResNet50) with our model that incorporates the localization task as well, our results outperform the ResNet50 on the same dataset.
Note that the method exhibited high sensitivity for COVID-19 pneumonia detection, thus proving its capability to detect COVID-19 pneumonia in addition to non-COVID-19 pneumonia.

To summarize, the results in Table \ref{table:detection_and_severity_results} show high performance on non-COVID-19-pneumonia detection ($Dataset\ 1$), and an even higher performance on COVID-19-pneumonia detection ($Dataset\ 2$), despite the fact that the network was not trained on COVID-19 images. Including COVID-19 images in the training dataset ($Dataset\ 3$) yielded even better performance, competitive with the state-of-the-art. The joint learning of detection and localization achieved higher detection results as compared to results from a system focusing on only one of the tasks.


\subsection{COVID-19 Severity Scoring and Follow-up}
\label{COVID-19 Severity Scoring Section}

\begin{table}[t]
\caption{Lung segmentation results reported for both the U-net based VGG-16 encoder method and the same method with additional abnormal datasets and augmentations. Ground truth masks were   generated using \cite{selvan2020lung}.}
\label{table:lung_segmentation_results}
\centering
\resizebox{0.80\columnwidth}{!}{
 \begin{tabular}{ccc} \toprule
 $Method$ & $Dice$ & $Jaccard$ \\ \midrule
LSNet & $0.89\pm0.07$ & $0.81\pm0.10$ \\ 
LSNet-Aug & $0.92\pm0.09$  & $0.86\pm0.11$ \\ \bottomrule
 \end{tabular}}
\end{table}

\begin{figure*}[!t]
  \begin{subfigure}[b]{0.5\textwidth}
    \includegraphics[width=\textwidth]{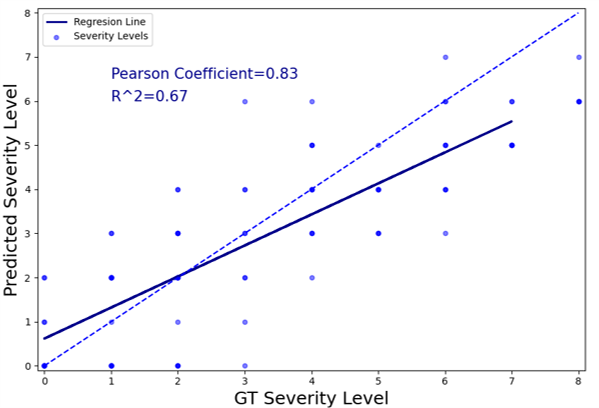}
    \caption{}
    \label{fig:severity_levels_regression}
  \end{subfigure}
  \begin{subfigure}[b]{0.5\textwidth}
    \includegraphics[width=\textwidth]{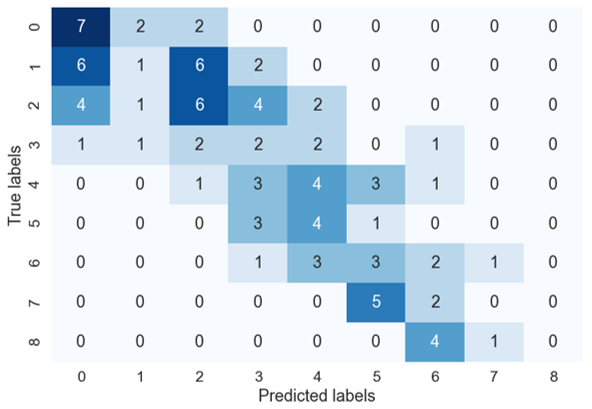}
    \caption{}
    \label{fig:confusion_matrix}
  \end{subfigure}
  \caption{(a) Scatter plot showing the relationship between the predicted and GT severity level. The dashed line corresponds to a perfect correlation and the solid blue line shows our linear regression model. (b) Confusion matrix showing the number of images that were scored with different combinations for severity scoring.}
\end{figure*}

\subsubsection{Lung Segmentation Evaluation}
Lung segmentation is essential to calculate an accurate severity score. We present a model for lung segmentation, dubbed $LSNet$, and suggest an improved model $LSNet-Aug$ that generalizes to images with severe infections such as COVID-19, by using various data augmentation techniques and including abnormal data sources.
The evaluation was run on 210 images provided in $Dataset\ 2$. The lung masks of these images were generated using the model described in \cite{selvan2020lung} as this achieved the most accurate segmentations. Therefore, we consider Selvan's method as our reference, and compared it to our lung segmentation models.
The results were evaluated using the Dice and Jaccard coefficient. Table \ref{table:lung_segmentation_results} shows an improvement in both metrics for lung segmentation after adding the augmentations and the datasets during training.

\subsubsection{Quantitative Analysis} 
$Dataset\ 2$ includes a cohort of 94 PA CXR images that are assigned a severity score of $[0/1/2/3/4]$, indicating the extent of ground glass opacity or consolidation in each lung (right and left lung). The images were labeled by three experts, based on score strategy adapted from \cite{wong2020frequency}. The opacity extent was scored as follows: $0 =$ no involvement; $1 = <25\%$ involvement; $2 = 25-50\%$ involvement; $3 = 50-
75\%$ involvement; $4 = >75\%$ involvement. The total extent score  in both lungs ranged from 0 to 8. To compare our results to the GT scores, we computed the pneumonia ratio for each lung. We divided the pneumonia ratio into four levels using the same GT  criterion, and the total score was summed for both lungs. The severity scoring is evaluated using a correlation coefficient of the fitted model between the predicted and the GT scores.

Fig. \ref{fig:severity_levels_regression} shows the predicted severity scores against the GT scores from the 94 patient cohort. The correlation coefficient of the fitted model was $0.83$ and $R^2=0.67$. These results exceed the reported severity estimation reported in \cite{cohen2020predicting}, which was tested on a subset of the same dataset. 
The confusion matrix in Fig. \ref{fig:confusion_matrix} shows larger confusion between close severity scores such as the low levels [0, 1, 2]. Even though the high severity level images are slightly underestimated, none were scored as a mild condition stage and vice-versa.
Given the high inter-rater variability, our plots show satisfactory agreement.

Rows $6-9$ in Table \ref{table:detection_and_severity_results} 
presents the severity scoring performance using the different development phases of our method and demonstrates the improvement of the severity measure for each component; Using the basic model for lung segmentation, $LSNet$, and the localization model after the second stage of training for the pneumonia localization ($DLNet-2$), performance was higher than using the localization model after the first stage of training.
The accuracy of the severity score vs. the GT further improved when using the advanced model for lung segmentation ($LSNet-Aug$). 
These results exceed the reported severity estimation reported in \cite{cohen2020predicting}, which was tested on a subset of the same dataset.

\begin{figure}[!t]
  \centering
  \includegraphics[width=0.49\textwidth]{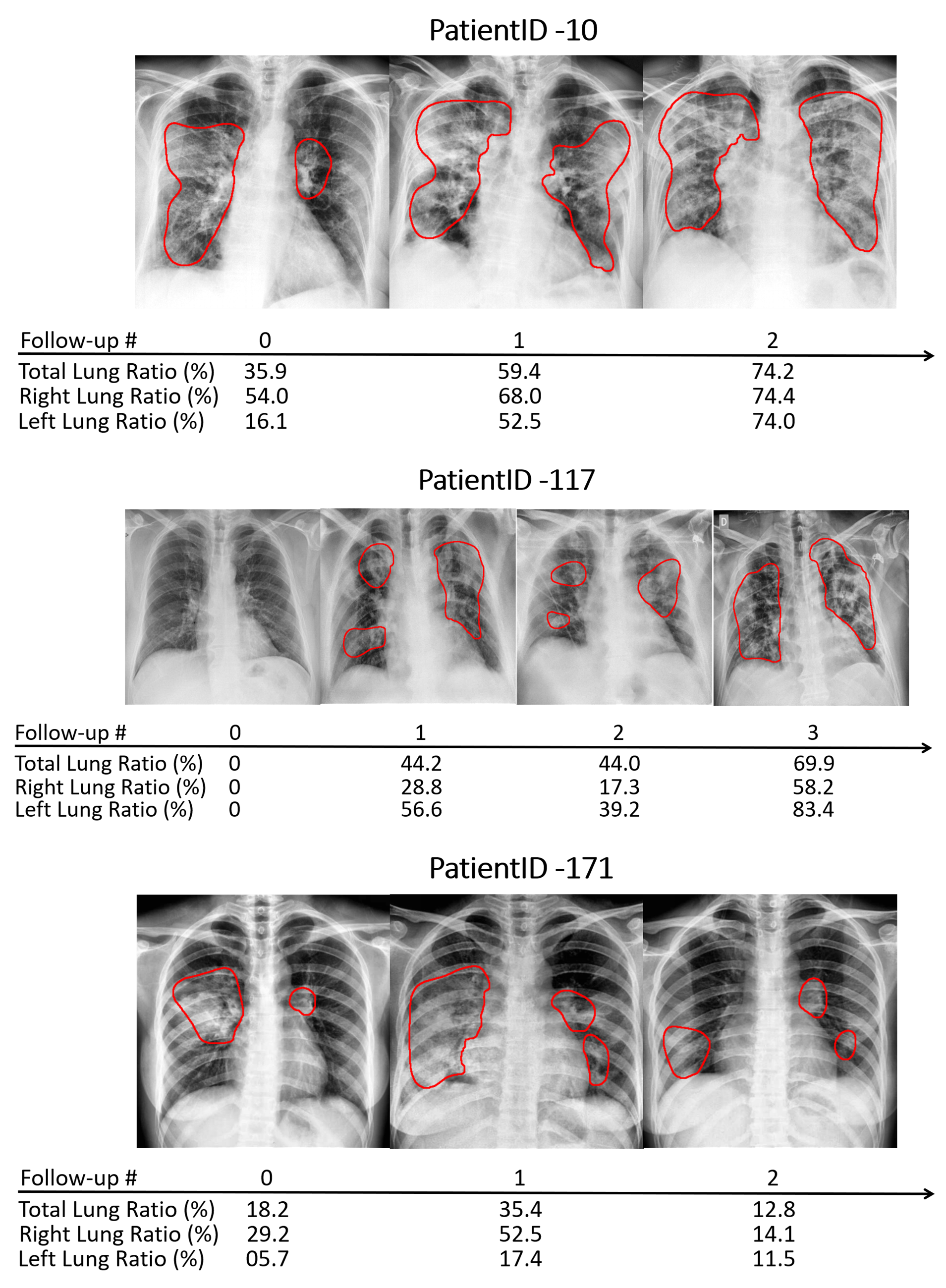}
  \caption{Example of patient monitoring over time in three patients using the pneumonia ratio.}
  \label{fig:patients_follow_up}
\end{figure}

\subsubsection{Qualitative Analysis} 
To estimate the progression and severity of pneumonia in COVID-19 patients, we explored the pneumonia ratio for patients from $Dataset\ 2$, scanned at multiple time points. The intervals between patients' two subsequent time points were inconsistent and ranged from 1-8 days. Thus the analysis did not rely on time intervals. We provide a qualitative analysis over time for three selected patients. Fig. \ref{fig:patients_follow_up} shows the CXR scans of these patients, superimposed with red contours indicating the predicted regions of pneumonia. The pneumonia ratio indicates the severity of pneumonia in these patients in percentages out of the lung field, right lung and left lung. In patient 10, the pneumonia ratio shows evidence of disease deterioration over time. In patient 117, there was a substantial increase followed by a period without major change and then another increase in disease severity. In contrast, for  patient 171 the ratio indicated recovery from the disease following a substantial increase in level of infection. 

\subsection{CT-Xray duality for patient monitoring validation }
To further validate the method, a quantitative analysis based on the strategy described in \ref{subsection:DRR} was performed. The $DeepDRR$ framework was applied to 9 patients with severe disease as indicated by their measured infiltration volume in CT.
The patients were scanned at Wenzhou hospital in China and were diagnosed with COVID-19 with the RT-PCR test. Each patient had a chest CT scan (slice thickness, $\{1,1.5\}\ mm$) at one or multiple time points (up to 4). The first CT scan
was obtained 1-4 days after the manifestation of the first signs of the virus (fever, cough) and the intervals between each two points ranged from 3 to 10 days. 

\begin{figure}[!t]
  \centering
  \includegraphics[width=0.5\textwidth]{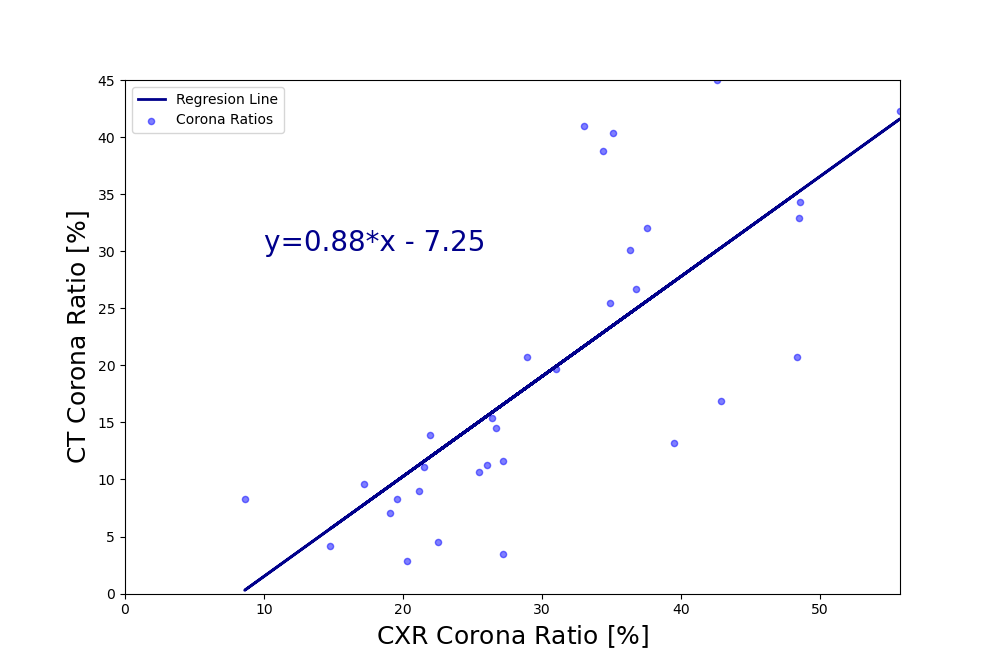}
  \caption{Linear regression model depicting the relationship of the pneumonia ratio on DRRs vs. the ratio calculated on the CT volume.}
  \label{fig:Linear_regression}
\end{figure}

After generating the DRRs, we applied our pneumonia detection and localization method (without re-training the model), and the pneumonia ratio was computed for each patient's generated Xray. The ratio of the detected infection in the lungs was also computed from the CT volume, as done in \cite{gozes2020rapid, gozes2020coronavirus}. 
A brief summary of the CT-based solution is the following: Following a lung segmentation module (based on \cite{frid2018improving}), a ResNet50 is used to classify the lung regions of each CT slice. For each positive (COVID-19) slice, a Grad-CAM procedure \cite{selvaraju2017grad} is utilized to generate a fine-grained localization map. These localization maps are used to calculate the ''Corona Score" by summation of all the pixels above a predetermined threshold. The $AUC$ of COVID-19 detection of this method was 0.99 with 0.94 sensitivity and 0.98 specificity, which makes this method a gold standard compared to Xray. A linear regression model was fitted to the CT and CXR pneumonia ratio values, as shown in  Fig. \ref{fig:Linear_regression}. 
The correlation coefficient between the two methods was $0.74$ ($p < 0.001$), where the slope of the line was 0.87 and the intercept with the y-axis was -7.2, thus indicating overall agreement.
\\
Fig. \ref{fig:DRR_follow_up} shows the pneumonia ratios extracted from the DRRs and the ratios computed on CT volumes, for each time point per patient. Our aim was to compare disease progression trends using the two modalities. 
We observed that in most cases the trend of the regression lines was similar; i.e., when the CXR ratio increased, so did the CT ratio, and vice-versa. The agreement between the two lines was quantified as follows: for each time point, if the quotient of the current time point to the previous point was greater than one, the sample was assigned a label of 1, otherwise 0. Using this definition, if the two lines agree in terms of their labels, we consider this to be a true prediction and the reverse (see agreement in green in Fig. \ref{fig:DRR_follow_up}). Overall, we computed an accuracy of 0.87 between the CT and Xray trends.
The differences in ratio values between the CT and CXR are worth noting. These are expected since the former was computed over the 3D volume, and the latter on a 2D image. We expected to see a dominant infiltration in CXR when the disease reached an advanced stage in CT, as depicted in Fig. \ref{fig:DRR_follow_up}. When CT is severe, the CXR ratio goes in the same direction, but after 3 or 4 time points (shaded points), the pneumonia infiltration decreases and the ratios in the CXR start to be less accurate. 

\section{Discussion}
\label{section:discussion}
The recent outbreak of COVID-19 has increased the need for automatic diagnosis and prognosis of COVID-19 pneumonia infections in CXR images. This includes the automatic follow-up of coronavirus patients to monitor their condition and the progression of the disease.
In this work, we present an end-to-end solution for COVID-19 pneumonia detection, localization, and severity scoring in CXR.
\begin{figure}[!t]
  \centering
  \includegraphics[width=0.5\textwidth]{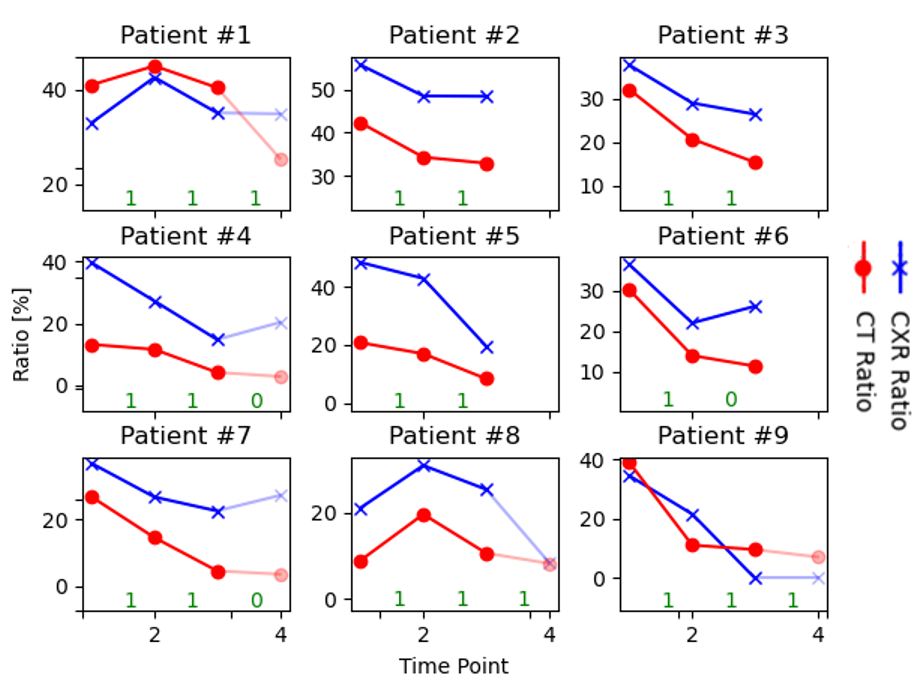}
  \caption{Comparison of patient monitoring using the pneumonia ratios computed from CT volumes and the corresponding DRRs. The numbers in green (on the x-axis) represent the agreement between change trends of CXR and CT ratios that were used to calculate the accuracy of change.
  Interval between time points 3-4 is shaded out to reflect mild disease states based on the CT.
  }
  \label{fig:DRR_follow_up}
\end{figure}

The severity of pneumonia is directly associated with its extent in the lungs; thus, an accurate segmentation of the regions infected with pneumonia is crucial. In this paper we present a dual-stage training scheme in a detection and localization network, to accurately segment infected pneumonia regions from inaccurate GT bounding boxes. To achieve reliable and accurate segmentation, we developed a weakly supervised method that exploits bounding box information and refines it in two stages.

Unlike previous works that have used Grad-CAM to provide clinically interpretable saliency maps \cite{rajaraman2020iteratively, lv2020cascade, oh2020deep} , we output the accurate localization directly from the network and prove its accuracy through the pneumonia severity scoring. The localization maps provided by our network demonstrate our model's ability to learn features that are specific to the disease, thereby showing that the calculations were not dataset-biased.

Several other works have attempted to solve the problem of detecting COVID-19 in CXR images \cite{wang2020covid, apostolopoulos2020covid, zhang2020covid}. Most networks have been trained and tested on COVID-19 patients with highly imbalanced labels from distinct datasets, on relatively small testing sets. This raises the concern that the network solutions may be dataset-biased, and not as robust as desired \cite{maguolo2020critic, tartaglione2020unveiling}. In order to assure the robustness of our solution we took special care to train the network on a single dataset that included non-COVID-19 pneumonia.
In the inference phase, we tested the method on a larger dataset including COVID-19 patients from an external public dataset and achieved high performance in these cases. Including COVID-19 cases in the training phase improved the network's results and yielded performance values comparable to the state-of-the-art with AUC, sensitivity and specificity values of 0.98, 0.92 and 0.97 respectively. 

Lung segmentation is less accurate in pathological lungs, specifically in severe conditions of pneumonia. We addressed this issue by using unconventional augmentations in the training process, including synthesizing pathological lungs from normal lung cases and adding blobs to the images along with a gamma correction. By applying these augmentations we were able to improve the segmentation results considerably and overall enhance the network performance.

A measure of the relative pneumonia region to the total lung region was found to strongly correlate with the disease severity score estimation.
In Table \ref{table:detection_and_severity_results} we presented an analysis of the effect of each training stage and the improved lung segmentation on the performance of the severity scoring against the GT labels, and show the contribution of each development step over previous works \cite{cohen2020predicting}, with a correlation coefficient of 0.83 and R2 = 0.67.
In future work, we plan to consider merging both lung segmentation and pneumonia detection into one architecture.

To validate patient-specific disease progression profiles, we need a disease score per time-point as the GT. The lack of such data prompted us to search for an alternative: we propose a novel validation scheme of synthesizing Xray (DRR) from CT using the $DeepDRR$ AI-based technique to show a proof of concept for patient monitoring in CXR. We used the proposed CT-Xray duality for longitudinal comparison to assess the disease state and trends in the severity of COVID-19 patients over time, which yielded an overall accuracy of 0.87 between the CT and Xray trends.
In our analysis, we utilized cases of severe illness. We focus on these cases due to the lower sensitivity of the Xray in comparison to the CT in detecting pneumonia for mild scenarios. This is exemplified in patient \#9 in Fig. \ref{fig:DRR_follow_up}, where the graph shows a ratio of 0 in CXR (indicating that the patient is negative for pneumonia), whereas the CT shows a positive ratio. Therefore, the pneumonia ratio is more accurate in monitoring patients at an advanced stage of the disease. 


\section{Conclusion}
We presented a model that simultaneously detects and localizes the region of pneumonia and assesses its extent in the lungs.  
We suggest dual-stage training that leverages the weak annotations of bounding boxes in order to output an accurate segmentation of pneumonia in COVID-19 patients.
An improvement on a previous lung segmentation method is described using unconventional additions that enhance the results of lung segmentation on diseased lungs.
The pneumonia and lung segmentation are exploited to quantify a pneumonia ratio which indicates the extent of pneumonia in the lungs. Additional exploration in the CT-Xray coupling is described to validate the ability of our method to monitor patients over time. Findings point to the utility of AI for COVID-19 pneumonia quantification, severity scoring and patient monitoring.

\printbibliography

\end{document}